\documentclass[american,english,aps,manuscript]{revtex4}
\usepackage[T1]{fontenc}
\usepackage[latin9]{inputenc}
\usepackage{float}
\usepackage{amssymb}
\usepackage{graphicx}
\usepackage{esint}

\makeatletter
\@ifundefined{textcolor}{}
{%
 \definecolor{BLACK}{gray}{0}
 \definecolor{WHITE}{gray}{1}
 \definecolor{RED}{rgb}{1,0,0}
 \definecolor{GREEN}{rgb}{0,1,0}
 \definecolor{BLUE}{rgb}{0,0,1}
 \definecolor{CYAN}{cmyk}{1,0,0,0}
 \definecolor{MAGENTA}{cmyk}{0,1,0,0}
 \definecolor{YELLOW}{cmyk}{0,0,1,0}
}

\makeatother

\usepackage{babel}
\begin{document}
\selectlanguage{american}%

\title{Collapses and revivals of matter waves}

\selectlanguage{english}%

\author{Hagar Veksler and Shmuel Fishman}

\address{Physics Department, Technion- Israel Institute of Technology, Haifa
32000, Israel}
\selectlanguage{american}%
\begin{abstract}
Quantum collapses and revivals are fascinating manifestations of interference.
Of particular interest in recent years are macroscopic quantum interference
effects in Bose-Einstein condensates. In this communication such effects
will be studied for the two site Bose-Hubbard model that is a standard
model for exploration of Bose-Einstein condensates. An analytic expression
that is valid in the weak coupling limit for the difference in the
occupation of the two sites is developed and tested numerically. It
describes correctly the collapses and revivals. Moreover, it is demonstrated
that a calculation to the first order in the interparticle interaction
is required for the prediction of the collapse and revival times while
the second order is required for the evaluation of the shape of the
revival peaks. We believe that the result is relevant for a variety
of situations where collapses and revivals are found.
\end{abstract}
\maketitle
\selectlanguage{english}%
Collapses and revivals are fascinating wave phenomena. By a collapse
one means that a pattern or an expectation value that is initially
pronounced, practically vanishes after some time, and by revival we
mean that at a latter time the pattern nearly returns to its initial
value. In optics such an effect is the Talbot effect discovered nearly
200 years ago \cite{Talbot,Rayleigh} (see also \cite{Berry_box}).
A related phenomenon is the {}``Quantum carpet'' \cite{Berry&Sclich,Berry_box}.
\foreignlanguage{american}{For the Jaynes-Cummings (JC) model \cite{Jaynes_model}
that is central in quantum optics, an expression exhibiting collapses
and revivals was found analytically \cite{Eberly} and it is of the
form reminiscent of the one found in the present work for a model
of interacting bosons in a regime of parameters that is of experimental
relevance.} Collapses and revivals can be found in many situations.
A generic picture is outlined in \cite{of1_pit,PS_review}.

In recent years, matter waves such as for Bose-Einstein condensates
(BECs) and other systems involving cold atoms were extensively studied
\cite{Pita_book,PS_review,Pethick@Smith}. Collapses and revivals
were observed in experiments where a BEC was confined to a lattice.
The interference pattern of the matter wave field originating in various
lattice sites showed collapses and revivals as a function of time
\cite{Bloch_colapse,Bloch_ex}. These were found also experimentally
for other condensates \cite{Deepak,Deepak2}. Collapses were found
in numerical calculations in the framework of a theoretical model
similar to the model we use here \cite{Milburn}. These studies are
related to the double well problem. This is a system defined by a
potential with two minima of equal depth separated by a barrier. The
potential is infinite at infinity. In the context of the present paper,
a large number of bosons $N$ is trapped in this potential. It was
studied experimentally \cite{Obert_exp,jeff,shin,dorond}. In particular,
the Bosonic Josephson effect attracted much interest \cite{Obert_exp,jeff},
since it is a clear manifestation of macroscopic quantum coherence.
It encourages theoretical exploration of this and related systems
\cite{Milburn,Doron,Smerzi,Spekkens&Sipe,ReinhardtPRL,ODell,ODell2}.
The double well was explored in detail \cite{Ofir_BH,Ofir_DW,OfirDW}.
In particular, it was shown that collapses take place \cite{Ofir_DW}
and agreement with the result of the Bose-Hubbard (BH) model was found
for inter-particle interactions that are very weak \cite{Ofir_BH}.
Collapses and revivals were found theoretically for interacting bosons
in a harmonic well \cite{Lewenstein,Wright,wright&walls,castin&dalibard,You},
for wave packets in harmonic wells with small nonlinearities \cite{Mark,PS_review,of1_pit},
for dynamics of atoms on optical lattices \cite{Fisher3,Fisher4}
and in experiments on Rydberg atoms \cite{of2,of3}.

The BH model where bosons can occupy only two sites \cite{Assa} was
studied extensively numerically and analytically \cite{Milburn,Doron,Ofir_DW,Ofir_BH,Kuklov4,Strunz}.
It is an approximation of the double well model in the limit where
the inter-particle interaction is sufficiently weak so that only the
two lowest levels of the double well are occupied. For the double
well, collapses were found in exact numerical calculations \cite{Ofir_DW}
and it was demonstrated that some of the results are similar to those
found for the BH model. The static properties of weakly interacting
BECs are often described by the Gross-Pitaevskii Equation (GPE). This
is not the case for the dynamics as was demonstrated for the double
well potential \cite{Smerzi}. In particular, it does not reproduce
neither collapses nor revivals. The BH is not only an approximation
to the double well, it can also be realized where the two sites are
the degenerate ground states of particles in a harmonic well \cite{Obert_phase}.

Analytical results on the time-evolution of interacting many-particle
systems are relatively rare, to the best of our knowledge there is
only one approximate analytical formula for a collapsing and reviving
quantity for a specific model of interacting bosons that was derived
in a controlled way \cite{Strunz}. This formula is not an explicit
expression in terms of the parameters of the model. The main result
of this communication is an explicit expression for the difference
in occupation of the two sites of the BH model for weak interaction
$u$ between the particles and large number of particles $N.$ It
is reminiscent of the one found for the JC model where the physics
is completely different \cite{Eberly}. In particular, we find that
in the first order in $\frac{1}{N}$ revival and collapse times are
found correctly using just the first order in $u$ while the shape
of the reviving peaks requires the second order in $u$. A detailed
version will be published \cite{long}. It is of experimental relevance
as demonstrated by various experiments and of conceptual importance
for understanding of coherence. The reason is that for the collapse
we discuss no information is lost and after some time a revival takes
place. It should be distinguished from collapses that often take place
in experiments where information is lost.

The calculations in this paper will be preformed in the framework
of the two site BH model. It is defined by the Hamiltonian
\begin{equation}
H_{BH}=-J\left(a_{L}^{\dagger}a_{R}+a_{R}^{\dagger}a_{L}\right)+U\left[n_{L}\left(n_{L}-1\right)+n_{R}\left(n_{R}-1\right)\right].\label{eq:H_BH_n}
\end{equation}
The sites are denoted by $L$ (Left) and $R$ (Right). The creation
and annihilation operators on the sites are $a_{L}^{\dagger},a_{R}^{\dagger}$
and $a_{L},a_{R}$. The number operators for the two sites are $n_{L}=a_{L}^{\dagger}a_{L}$
and $n_{R}=a_{R}^{\dagger}a_{R}$. The commutation relations are $\left[a_{L},a_{L}^{\dagger}\right]=1$,
$\left[a_{R},a_{R}^{\dagger}\right]=1$,\foreignlanguage{american}{
and the units are such that $\hbar=\frac{1}{N}$. It is assumed that
the site energies on the two sites are identical. The total number
of particles $n_{L}+n_{R}=N$ is conserved. The first term in (\ref{eq:H_BH_n})
represents the hopping between the two sites while the second one
is the energy of the interparticle interaction. The BH Hamiltonian
(\ref{eq:H_BH_n}) can be written up to multiplicative and additive
constants as
\begin{equation}
H=-S_{x}+uS_{z}^{2}\label{eq:H_sx_sz}
\end{equation}
where $u\equiv\frac{UN}{J}$ and $\overrightarrow{S}=\left(S_{x},S_{y},S_{z}\right)$.
The components of $\overrightarrow{S}$ are $S_{x}=\frac{1}{2N}\left(a_{R}^{\dagger}a_{L}+a_{L}^{\dagger}a_{R}\right)$,
$S_{y}=\frac{i}{2N}\left(a_{R}^{\dagger}a_{L}-a_{L}^{\dagger}a_{R}\right)$,
$S_{z}=\frac{1}{2N}\left(a_{L}^{\dagger}a_{L}-a_{R}^{\dagger}a_{R}\right)=\frac{1}{2N}\left(n_{L}-n_{R}\right)$.
They satisfy the standard commutation relations of the angular momentum,
with $\hbar$ replaced by $\frac{1}{N}$. The relation between $H_{BH}$
and $H$ is
\begin{equation}
H_{BH}=2JNH+\frac{1}{2}UN^{2}-NU.
\end{equation}
For large $N$ the Semiclassical analysis of the dynamics generated
by $H$ is useful \cite{Milburn,Doron,dorona,doronb}. The Josephson
regime $1<u<N^{2}$ was extensively studied. Here we confine ourselves
to the Rabi regime $u<1$. First, we note that $S^{2}=\frac{1}{2N}\left(\frac{N}{2}+1\right)$
is a constant of motion. The classical dynamics of the vector $\vec{S}$
is the motion on the Bloch sphere of radius $\frac{1}{2}$. For $u<1$
it is just a motion with an angle $\varphi$ around the $x$ axis,
that is $\vec{S}=\left(S_{x},S_{\perp}\cos\varphi,S_{\perp}\sin\varphi\right)$
with $S_{\perp}=\sqrt{1-S_{x}^{2}}$. The Hamiltonian (\ref{eq:H_sx_sz})
takes the form
\begin{equation}
H=-S_{x}+u\left(\frac{1}{4}-S_{x}^{2}\right)\sin^{2}\varphi,\label{eq:H_sx_fi}
\end{equation}
where $S_{x}$ and $\varphi$ are conjugate variables. For $u=0$
the angular frequency is $\dot{\varphi}=\frac{\partial H}{\partial S_{x}}=-1$,
therefore, in these units the normalized difference in the occupation
between the Left and Right sites is proportional to $S_{z}$ and to
$\sin t$.}

\selectlanguage{american}%
As a result of the contribution of the second term in (\ref{eq:H_sx_fi}),
$\dot{\varphi}$ is not a constant but exhibits small variations.
Classically, the difference between the occupation of the Left and
Right sites oscillates with constant amplitude and period. Since $\varphi$
and $S_{x}$ are not angle-action variables, $\dot{\varphi}$ and
$S_{x}$ vary with time. The standard transformation to angle-action
variables $\left(I,\widetilde{\varphi}\right)$ is $I=\frac{1}{2\pi}\int_{0}^{2\pi}S_{x}d\varphi$
and $\dot{\widetilde{\varphi}}=\frac{\partial H}{\partial I}$. To
order $u$, the Hamiltonian is 
\begin{equation}
H\approx-I+\frac{1}{8}u-\frac{1}{2}uI^{2}.\label{eq:H_I}
\end{equation}
The action variable is quantized \cite{Tabor} so that
\begin{equation}
I_{n}=\frac{n}{N}
\end{equation}
where $n=-\frac{N}{2},...,\frac{N}{2}$ are integers. Note that $\dot{\varphi}=-\frac{\partial H}{\partial S_{x}}\approx-1$
for small $u$ and therefore $\dot{\varphi}$ never vanishes. Consequently
the Maslov index vanishes. Hence, the spectrum of the Hamiltonian
(\ref{eq:H_sx_sz}) is
\begin{equation}
E_{n}^{\left(1\right)}\approx-\frac{n}{N}+\frac{1}{8}u-\frac{1}{2N^{2}}un^{2}.\label{eq:En}
\end{equation}
The corresponding spectrum of the BH Hamiltonian is
\begin{equation}
E_{n}^{\left(BH1\right)}=2JNE_{n}^{\left(1\right)}+\frac{1}{2}UN^{2}-NU\approx2J\left(-n+\frac{3}{8}uN-\frac{1}{2}u-\frac{1}{2N}un^{2}\right).\label{eq:En-1}
\end{equation}
The calculation was extended to the second order in $u$, resulting
in (for details see \cite{long})\foreignlanguage{english}{
\begin{equation}
E_{n}^{\left(BH2\right)}\approx2J\left(-n+\frac{3}{8}uN-\frac{1}{2}u-\frac{1}{2N}un^{2}-\frac{1}{16}u^{2}n+\frac{1}{4N^{2}}u^{2}n^{3}\right).\label{eq:E_BH2}
\end{equation}
We verified that the expression (\ref{eq:E_BH2}) can be obtained
in the framework of standard second order perturbation theory in $u$,
where in each term only the leading contribution in $\frac{1}{N}$
was kept. It is important to notice that our result is the contribution
up to the order $u^{2}$ in the semiclassical approximation. It holds
for $u<1$, while the standard perturbation theory requires $uN<1$.
Finally we compared the pertubative results for the spectrum with
the ones obtained from direct diagonalization of (\ref{eq:H_BH_n})
and found excellent agreement even for $u=2$. Such results were encountered
also in other situations \cite{Doron_cp4}.}

\selectlanguage{english}%
We turn now to calculate the time dependence of the normalized difference
in population between the left and right sites (denoted by $\Delta\left(t\right)$).
We start from a state where all particles are on the Left site, in
this state $\left\langle S_{z}\right\rangle =\frac{1}{2}$. Since
$u$ is small, it is convenient to expand this state in terms of eigenstates
of $S_{x}$, and then calculate the correction of second order in
$u$ \cite{long}. Such states are 
\begin{equation}
\left|n\right\rangle \equiv\frac{1}{\sqrt{\left(\frac{N}{2}+n\right)!\left(\frac{N}{2}-n\right)!}}\left(a_{+}^{\dagger}\right)^{\frac{N}{2}+n}\left(a_{-}^{\dagger}\right)^{\frac{N}{2}-n}\left|0\right\rangle .\label{eq:S_x_eigen}
\end{equation}
where $a_{\pm}^{\dagger}=\frac{1}{\sqrt{2}}\left(a_{L}^{\dagger}\pm a_{R}^{\dagger}\right)$
and $\left[a_{+},a_{-}\right]=0$, $\left[a_{+},a_{+}^{\dagger}\right]=1$,
$\left[a_{-},a_{-}^{\dagger}\right]=1$. The reason for (\ref{eq:S_x_eigen})
is that $S_{x}=\frac{1}{2N}\left(a_{+}^{\dagger}a_{+}-a_{-}^{\dagger}a_{-}\right)$.

The initial state is
\begin{equation}
\left|\psi\left(t=0\right)\right\rangle =\frac{1}{\sqrt{N!}}\left(a_{L}^{\dagger}\right)^{N}\left|0\right\rangle =\frac{1}{2^{N/2}\sqrt{N!}}\left(a_{+}^{\dagger}+a_{-}^{\dagger}\right)^{N}\left|0\right\rangle .
\end{equation}
 It is useful to expand $\left|\psi\left(t=0\right)\right\rangle $
in the basis of (\ref{eq:S_x_eigen}),
\begin{equation}
\left|\psi\left(t=0\right)\right\rangle =\sum_{n=-N/2}^{N/2}c_{n}\left|n\right\rangle ,\label{eq:i}
\end{equation}
 and for large $N$
\begin{equation}
c_{n}\approx\left(\frac{2}{\pi N}\right)^{\frac{1}{4}}e^{-\frac{n^{2}}{N}}.\label{eq:cn}
\end{equation}
We note that the normalized difference between the occupation of the
two sites is 
\begin{equation}
\Delta\left(t\right)=\left\langle \psi\left|S_{z}\right|\psi\right\rangle =\frac{1}{N}\mathrm{Re}\left\langle \psi\left|\widetilde{S}_{+}\right|\psi\right\rangle 
\end{equation}
where $\widetilde{S}_{+}\equiv N\left(S_{z}-iS_{y}\right)=\frac{1}{2}\left(a_{L}^{\dagger}+a_{R}^{\dagger}\right)\left(a_{L}-a_{R}\right)=a_{+}^{\dagger}a_{-}$.
In the basis $\left\{ \left|n\right\rangle \right\} $, $\widetilde{S}_{+}$
is a raising operator, therefore $\widetilde{S}_{+}\left|n\right\rangle \propto\left|n+1\right\rangle $.
For large $N$,
\begin{equation}
\left\langle \psi\left(t\right)\left|\widetilde{S}_{+}\right|\psi\left(t\right)\right\rangle =\frac{N}{2}\sum_{n=-N/2}^{N/2}c_{n}c_{n+1}e^{-i\left(E_{n}^{\left(BH2\right)}-E_{n+1}^{\left(BH2\right)}\right)t}.\label{eq:usm}
\end{equation}
Substitution of the energies (\ref{eq:E_BH2}) results for large $N$
in
\begin{equation}
\left\langle \psi\left(t\right)\left|\widetilde{S}_{+}\right|\psi\left(t\right)\right\rangle =\widetilde{S}e^{-i\phi t}
\end{equation}
where the phase $\phi$ will be specified at a later stage and 
\begin{equation}
\widetilde{S}=\frac{\sqrt{N}}{\sqrt{2\pi}}\sum_{n=-N/2}^{N/2}e^{-\frac{2n^{2}+2n+1}{N}}e^{-i\frac{J}{N}\left(2un-\frac{3}{2N}u^{2}n^{2}-\frac{3}{2N}u^{2}n\right)t}.\label{eq:18}
\end{equation}
Since $n$ is an integer, in first order in $u$, the envelope of
the sum (\ref{eq:usm}) is a periodic function of $t$ with period
(revival time) of 
\begin{equation}
T_{R}=\frac{\pi N}{uJ}.\label{eq:Tr}
\end{equation}
Around the $m$-th revival, we write $t=m\cdot T_{R}+\tau$ with $-\frac{1}{2}T_{R}<\tau<\frac{1}{2}T_{R}$
and (\ref{eq:18}) takes the form 
\begin{equation}
\begin{array}{ccc}
\widetilde{S}_{m} & = & \frac{\sqrt{N}}{\sqrt{2\pi}}\sum_{n=-N/2}^{N/2}e^{-\left(2n^{2}+2n+1\right)/N}e^{+i\frac{3}{2N^{2}}u^{2}n^{2}J\cdot\left(m\cdot T_{R}+\tau\right)-\frac{2}{N}iJun\tau}\end{array}.\label{eq:usm-1-2}
\end{equation}
Therefore, $\widetilde{S}=\sum_{m}\widetilde{S}_{m}$. Around each
revival the sum can be replaced by an integral since in the vicinity
of a revival $\frac{J}{N}u\tau+\frac{3}{N^{2}}u^{2}n\tau$ is small
(while $\frac{J}{N}unt$ is typically large). This will be discussed
in what follows. Doing the integral over $n$ for $-\frac{1}{2}T_{R}<\tau<\frac{1}{2}T_{R}$,
the final result is (for the detailed calculation see \cite{long}),
\begin{equation}
\Delta\left(t\right)=\left\langle \psi\left(t\right)\left|\widetilde{S}_{z}\right|\psi\left(t\right)\right\rangle =\frac{1}{2}\sum_{m}A_{m}\exp\left[\frac{-\left(t-mT_{R}+\frac{3m\pi}{2J}\right)^{2}}{\left(\Delta t_{R}^{m}\right)^{2}}\right]\cos\left(\phi_{1}-\phi t\right),\label{eq:S_z_good}
\end{equation}
where
\begin{equation}
\Delta t_{R}^{m}\approx\frac{\sqrt{2N\left(1+\frac{9}{16}u^{2}m^{2}\pi^{2}\right)}}{Ju},\label{eq:dtRm}
\end{equation}
\begin{equation}
A_{m}=\frac{e^{-\frac{u^{2}}{32}}}{\left[1+\frac{9}{16}u^{2}m^{2}\pi^{2}\right]^{1/4}}\exp\left[\frac{2+\frac{9}{2}u^{2}m^{2}\pi^{2}}{4N\left(1+\frac{9}{16}u^{2}m^{2}\pi^{2}\right)}\right],\label{eq:amp}
\end{equation}
\begin{equation}
\phi\approx J\left(2+\frac{1}{8}u^{2}+\frac{u}{N}\right).\label{eq:fi_main}
\end{equation}
and
\begin{equation}
\phi_{1}\approx\frac{u^{2}}{8}\left(2J\tau+\frac{3}{2}m\pi\right)+u\left(\frac{J\tau}{N}+\frac{3}{8}\left(m\cdot\pi+\frac{J}{N}u\tau\right)\right).\label{eq:fi1_main}
\end{equation}
This is a sequence of Gaussians of width $\Delta t_{R}^{m}$ of order
$\sqrt{N}$ with a separation $T_{R}$ of order $N$. For short times
$\left(m=0\right)$, the dynamics is described by 
\begin{equation}
\Delta\left(t\right)=\frac{1}{2}e^{-\frac{1}{2N}J^{2}u^{2}t^{2}}\cos\left(\phi t-\phi_{1}\right)\label{eq:Sz_t}
\end{equation}
and the collapse time is given by 
\begin{equation}
T_{c}=\frac{\sqrt{2N}}{Ju}.\label{eq:Tc}
\end{equation}
It is important to note that for small $m$ the width of the Gaussian
peaks is of the order $\frac{\sqrt{2N}}{Ju}$, therefore, for such
values of $m$, $\frac{J}{N}u\tau+\frac{3}{N^{2}}u^{2}n\left(mT_{R}+\tau\right)\leq\sqrt{\frac{2}{N}}+\frac{3}{JN}un\left(m\pi+\sqrt{\frac{2}{N}}\right)$
that is much smaller than $1$ ($\frac{u}{J}$ is small). Therefore
the sum (\ref{eq:usm-1-2}) can be approximated by an integral leading
to the relatively simple formula (\ref{eq:S_z_good}).

The evolution of the expectation of the normalized difference in occupation
of the two sites $\Delta\left(t\right)$ is the main result of the
present work. In Fig. 1 it is compared to exact results found by numerical
diagonalization of the Hamiltonian (\ref{eq:H_BH_n}), for $u=\frac{1}{2},\, N=100$
and $J=1$. We note remarkable agreement of the envelope with the
exact numerical result. The rapid oscillations, exhibit good agreement
for short times (Fig. 1(b)) but it deteriorates for longer times (Fig.
1(c)).

In Fig. 2 the evolution of the normalized difference in occupation
between the two sites is presented for $u=\frac{1}{20},\, N=50$ and
$J=1$. We note also the remarkable agreement between the analytical
and numerical results found for the envelope. The prediction for the
rapid oscillations agrees with the exact results for longer times
and more revivals than in Fig. 1.

\selectlanguage{american}%
\begin{figure}[H]
\selectlanguage{english}%
(a)\qquad{}\qquad{}\qquad{}\qquad{}\qquad{}\qquad{}\qquad{}\qquad{}\qquad{}\qquad{}\qquad{}\qquad{}(b)

\includegraphics[scale=0.45]{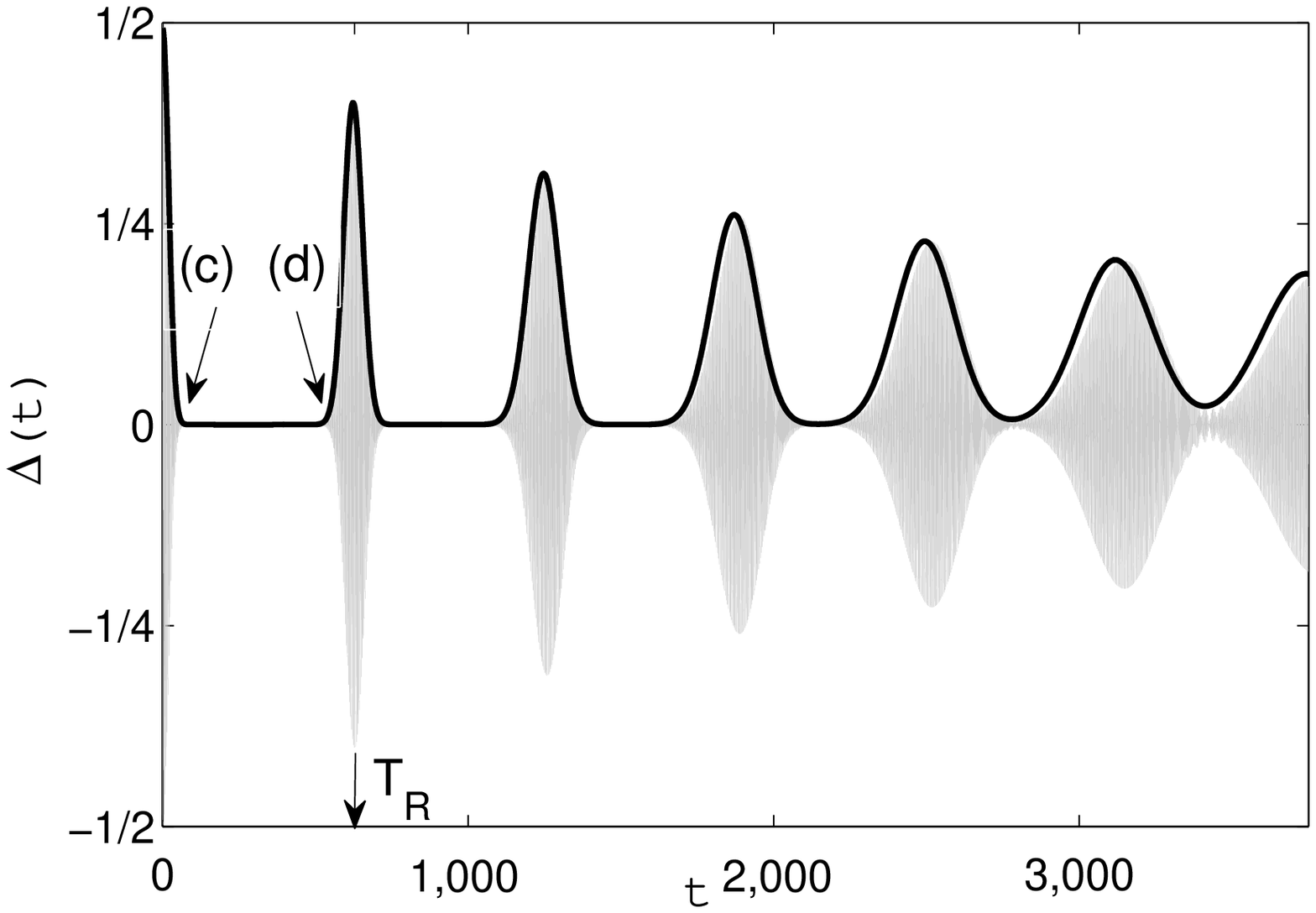}\includegraphics[scale=0.45]{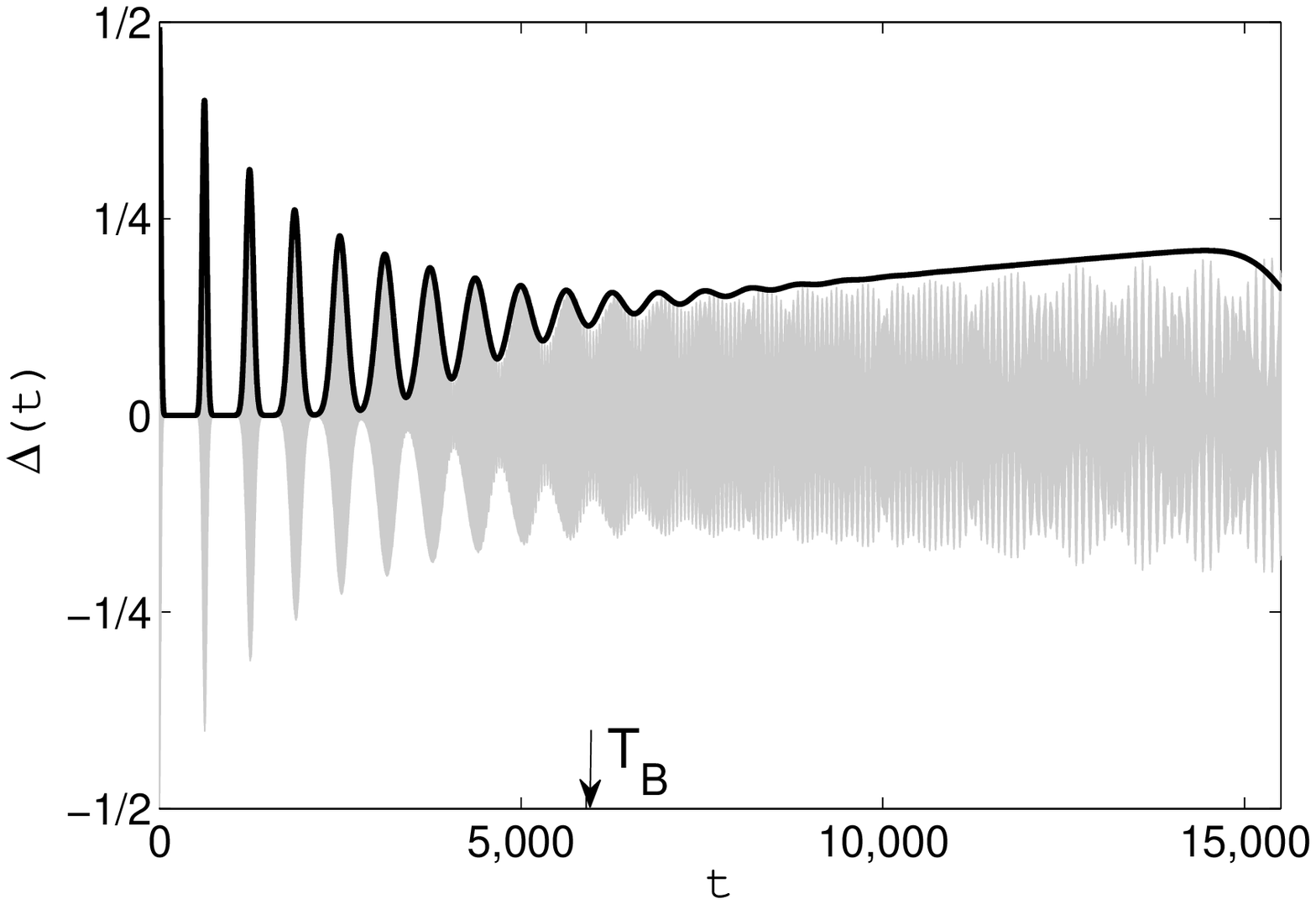}

(c)\qquad{}\qquad{}\qquad{}\qquad{}\qquad{}\qquad{}\qquad{}\qquad{}\qquad{}\qquad{}\qquad{}\qquad{}(d)

\includegraphics[scale=0.45]{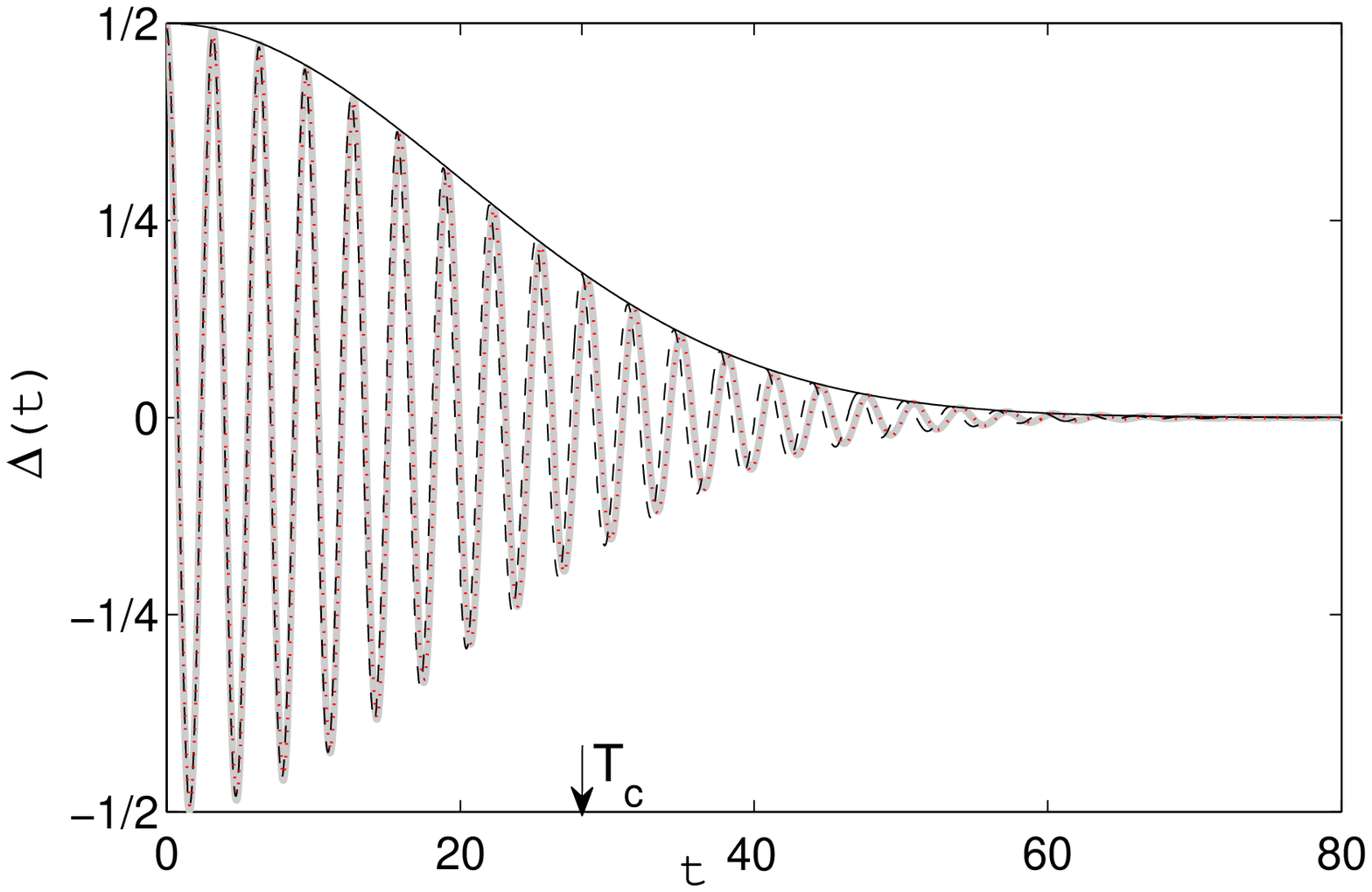}\includegraphics[scale=0.45]{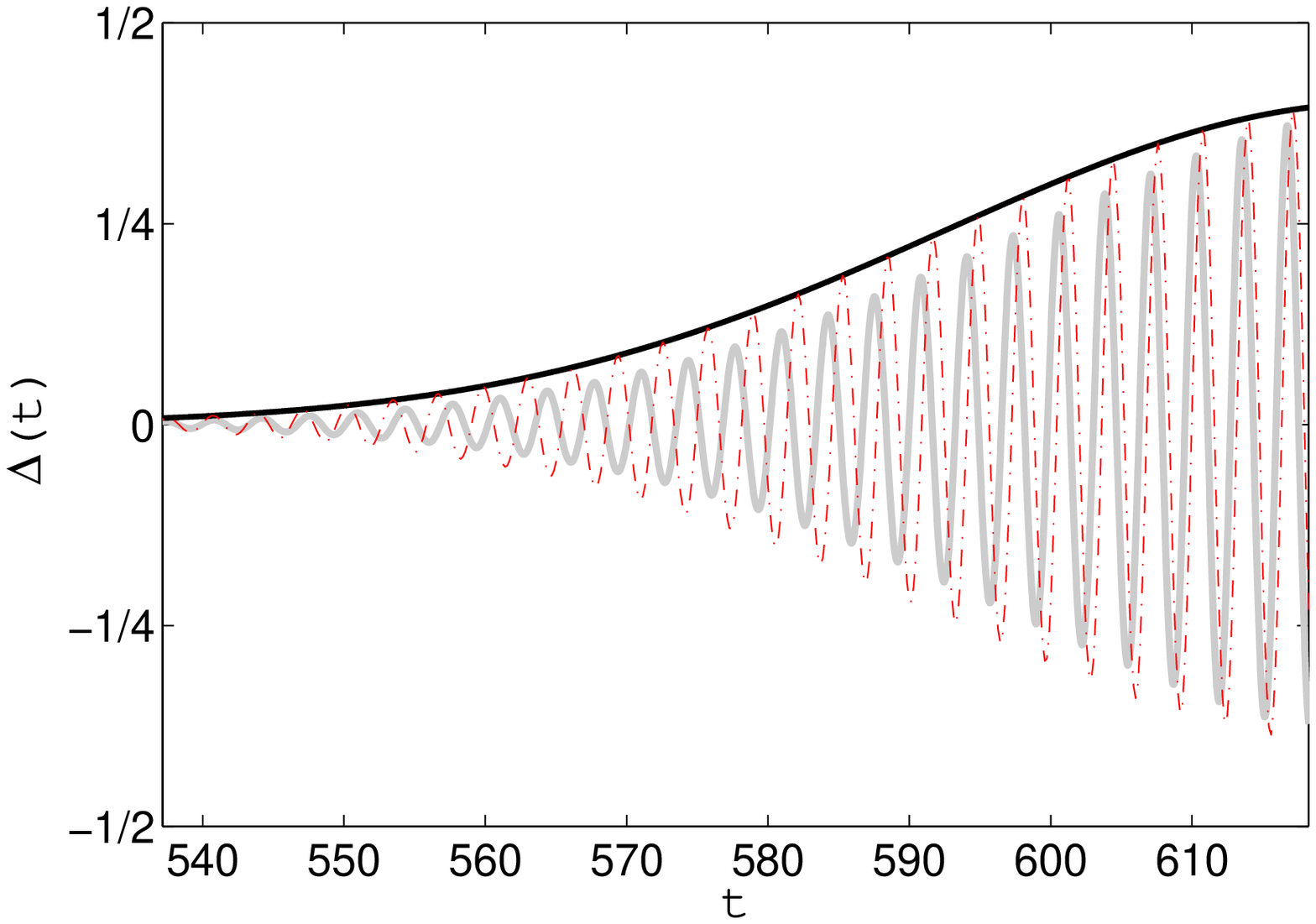}

\selectlanguage{american}%
\caption{\selectlanguage{english}%
(Color online) The normalized difference between the occupation of
the two sites $\Delta\left(t\right)$ for $J=1$, $N=100$ and $u=\frac{1}{2}$.
The light gray line represents the numerical result, obtained by diagonalizing
the Hamiltonian (\ref{eq:H_BH_n}). The black line represents the
envelope based on (\ref{eq:S_z_good}). (a) $\Delta\left(t\right)$
for the time regime $t<T_{B}$. The arrows show the time regimes which
are presented in (c) and (d). The time $T_{R}$ of (\ref{eq:Tr})
is marked. (b) Long time blurring. The time $T_{B}=m_{max}T_{R}$
where the revivals mix (see Eq. (\ref{eq:m_max})) is marked. (c)
Short time dynamics. The red dashed-dot line is given by (\ref{eq:S_z_good})
where $\phi$ and $\phi_{1}$ are given by (\ref{eq:fi_main}) and
(\ref{eq:fi1_main}). The dashed black line presents oscillations
with the unperturbed Rabi's frequency $2J$ (that is approximating
the phase $\phi t-\phi_{1}$ by $2Jt$) and $T_{c}$ of (\ref{eq:Tc})
is marked. (d) the same as (c) for a time interval near the revival
$m=1$, where the analytical result for the phase $\phi_{1}-\phi t$
no longer agrees with the result of exact numerical calculation.\selectlanguage{american}%
}
\end{figure}

\begin{figure}[H]
\selectlanguage{english}%
(a)\qquad{}\qquad{}\qquad{}\qquad{}\qquad{}\qquad{}\qquad{}\qquad{}\qquad{}\qquad{}(b)

\includegraphics[scale=0.4]{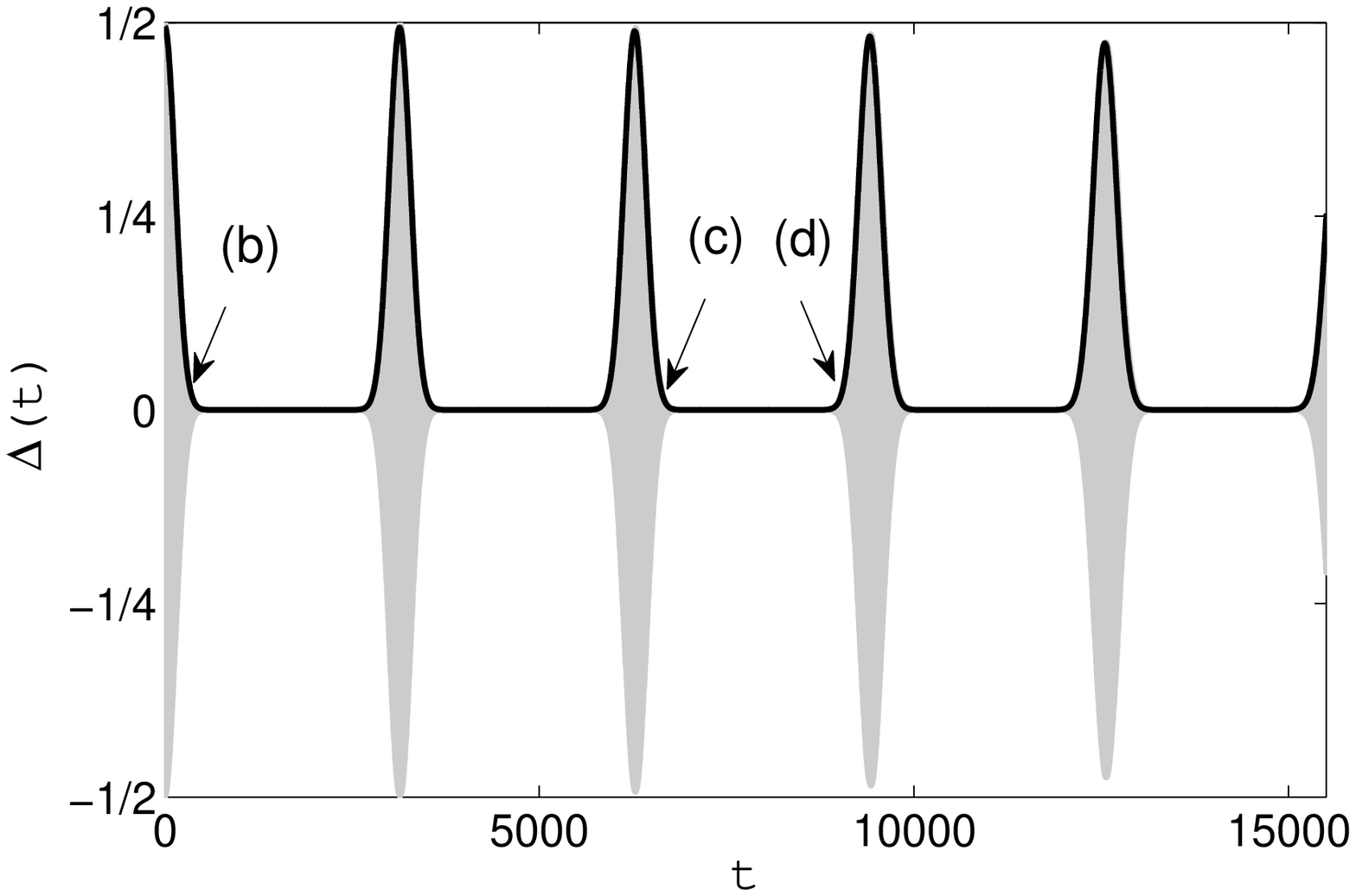}\includegraphics[scale=0.4]{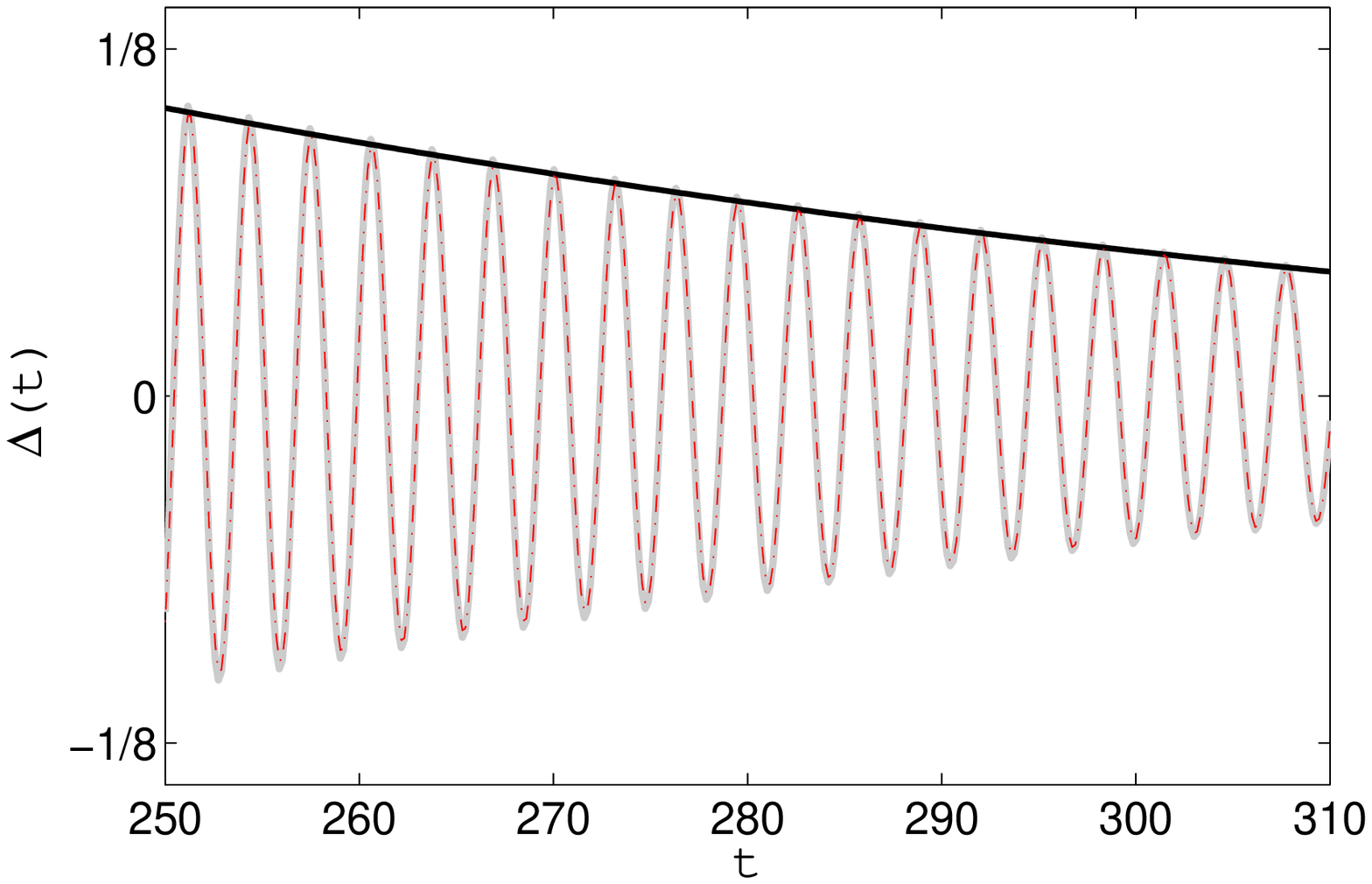}

(c)\qquad{}\qquad{}\qquad{}\qquad{}\qquad{}\qquad{}\qquad{}\qquad{}\qquad{}\qquad{}(d)

\includegraphics[scale=0.4]{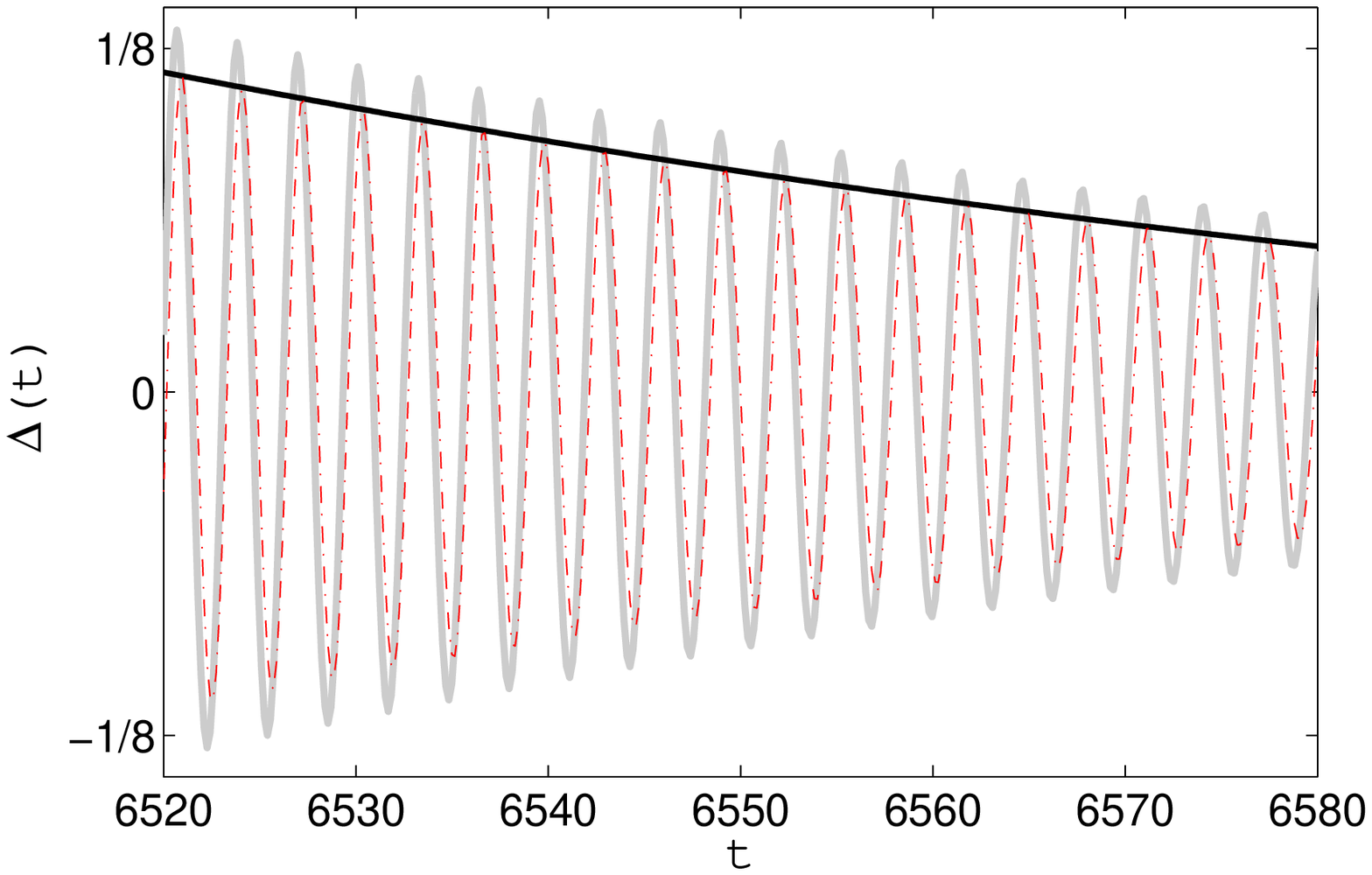}\includegraphics[scale=0.4]{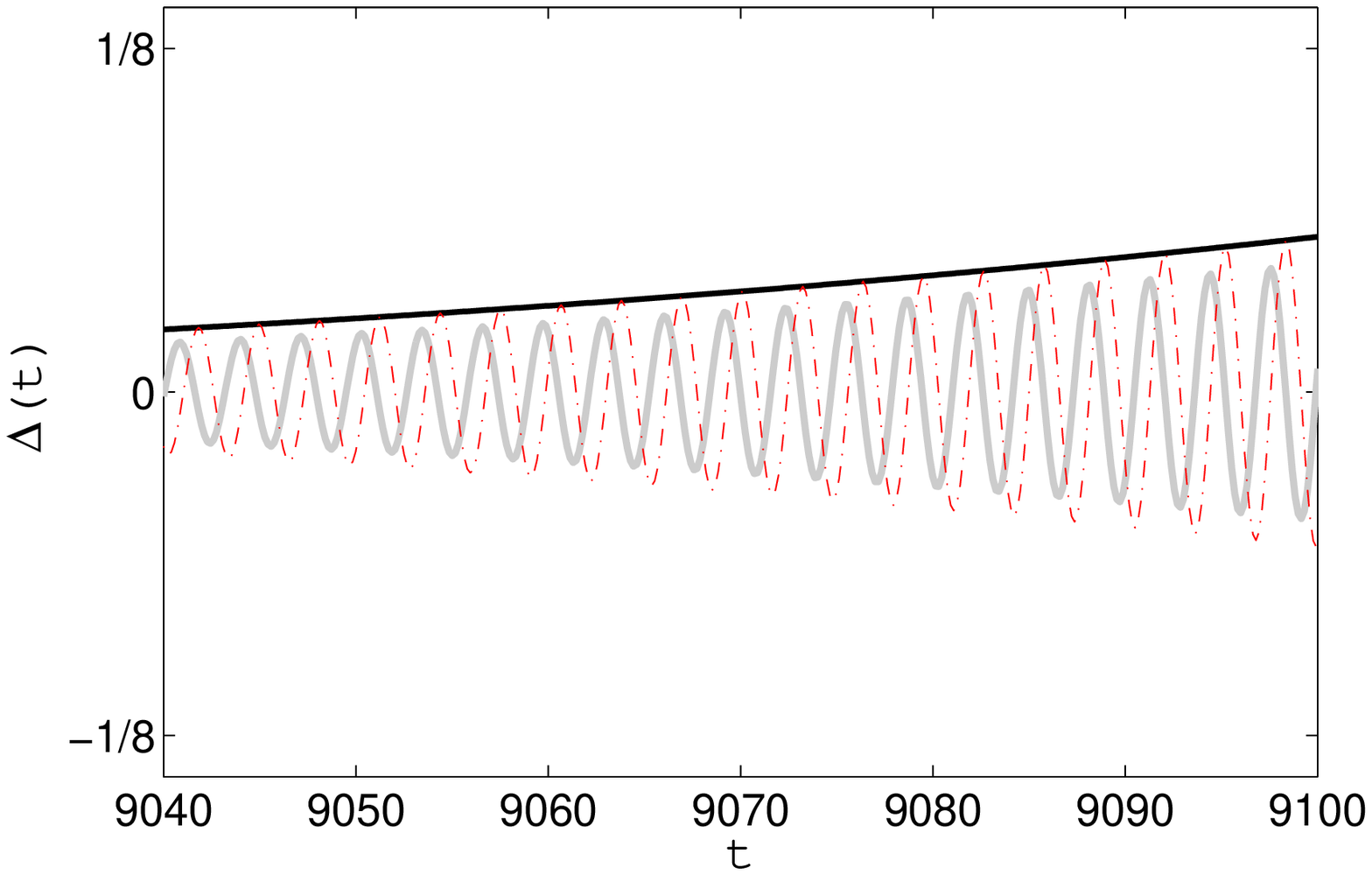}

\selectlanguage{american}%
\caption{\selectlanguage{english}%
(Color online) Similar to Fig. 1 but for $J=1$, $N=50$ and $u=\frac{1}{20}$.
(a) $\Delta\left(t\right)$ for a time regime $t<T_{B}$. The arrows
show the time regimes which are presented in (b)-(d). (b) Short time
dynamics. (c) the same as (b) for a time interval near the revival
$m=2$. (d) the same as (b) for a time interval near the revival $m=3$,
where the analytical result for the phase in (\ref{eq:S_z_good})
no longer agrees with the exact numerical calculation.\selectlanguage{american}%
}
\end{figure}

\selectlanguage{english}%
For small $m$, $\Delta t_{R}^{m}\ll T_{R}$. However, there is an
$m_{max}$ where the width $\Delta t_{R}^{m}$ is comparable to $T_{R}$
and then the revivals mix and our calculations are not valid. Defining
$m_{max}$ by $\Delta t_{R}^{m_{max}}=\frac{1}{2}T_{R}$, we estimate
\begin{equation}
m_{max}=\frac{\sqrt{2\left(\pi^{2}N-8\right)}}{3u\pi}.\label{eq:m_max}
\end{equation}
 We checked that indeed for $m>m_{max}$ the peaks mix and the picture
presented in Figs. 1(a) and 2(a) deteriorates. The calculation can
be extended to the case where initially both sites are occupied and
in some situations a formula similar to (\ref{eq:S_z_good}) is found.
If the initial occupation of the left site is given by $\cos^{2}\alpha$,
it exhibits revivals at times $T_{R}\left(1-\frac{3}{4}u\sin\left(2\alpha\right)\right)^{-1}$
where $T_{R}$ is given by (\ref{eq:Tr}) \cite{long}.

The main result of this paper is the analytic expression (\ref{eq:S_z_good})
for the normalized difference between the two sites occupation $\Delta\left(t\right)$
of the Bose-Hubbard model defined by (\ref{eq:H_BH_n}). It consists
of a sequence of Gaussian peaks, superimposed on a rapid oscillation.
Comparison between the approximate result and the exact numerical
calculation demonstrates that the result obtained indeed requires
the terms in order $u^{2}$ and $\frac{1}{N}$. The classical approximation
(\ref{eq:H_sx_fi}) reproduces correctly the rapid oscillations for
short times. Such a behavior is found also for the GPE in double well
\cite{Milburn,Smerzi}. Quantization is essential for the collapses
and revivals. The collapse and revival times are predicted correctly
by the first order in the interaction $u$, however for the width
of the peaks the order $u^{2}$ is required. The population difference
exhibits three time scales (superimposing the Rabi oscillations):
The collapse time $T_{c}$ (\ref{eq:Tc}), the revival time $T_{R}$
(\ref{eq:Tr}) and $T_{B}=m_{max}T_{R}$ (\ref{eq:m_max}) where the
revival picture is blurred. If initially both sites are occupied but
the imbalance is large, a similar picture emerges but the time scales
are different. We can see from (\ref{eq:amp}) that the amplitude
of the revival peaks decreases with time (see also Figs. 1 and 2).
This decrease is completely coherent. 

The result presented in this communication is a fascinating manifestation
of macroscopic quantum coherence. It is of great importance for distinguishing
collapses resulting of dephasing where quantum coherence is dumped
(and consequently the revivals are dumped as well) from the situation
presented in this work where quantum revivals are found. The knowledge
of the function $\Delta\left(t\right)$, and in particular the decrease
in the amplitude of the peaks that is completely coherent, can be
used to measure the rate of destruction of coherence in experiments.
The main result (\ref{eq:S_z_good}) can be used also for the comparison
between the Bose-Hubbard model and the double well problem \cite{Ofir_DW}.
It is interesting to note that (\ref{eq:S_z_good}) is very similar
to the result found in \cite{Eberly} for completely different physics.
We believe that the method of the calculation used here can be applied
to other physical situations as well, in particular in presence of
interactions.

This work resulted of a discussion with Doron Cohen on ref. \cite{Doron}.
We thank him for motivating this direction of research and many critical
discussions and communications. We thank also Ofir Alon, Or Alus and
I. Bloch for illuminating and informative discussions and communications.
The work was supported in part \foreignlanguage{american}{by the Israel
Science Foundation (ISF) grant number 1028/12, by the US-Israel Binational
Science Foundation (BSF) grant number 2010132 and by the Shlomo Kaplansky
academic chair.}


\begin{thebibliography}{10}

\bibitem{Talbot}
H.~Talbot,
\newblock Philos. Mag {\bf 9}, 401 (1836).

\bibitem{Rayleigh}
L.~Rayleigh,
\newblock Philos. Mag {\bf 11} (1881).

\bibitem{Berry_box}
M.~V. Berry,
\newblock J. Phys. A {\bf 29}, 6617 (1996).

\bibitem{Berry&Sclich}
M.~V. Berry, I.~Marzoli, and W.~Schleich,
\newblock Physics World , 39 (2001).

\bibitem{Jaynes_model}
E.~T. Jaynes and F.~W. Cummings,
\newblock Proc. Inst. Elect. Eng. {\bf 51}, 89 (1963).

\bibitem{Eberly}
J.~H. Eberly, N.~B. Narozhny, and J.~J. Sanchez-Mondragon,
\newblock Phys. Rev. Lett. {\bf 44}, 1323 (1980).

\bibitem{of1_pit}
L.~P. Pitaevskii,
\newblock Phys. Lett. A {\bf 229}, 406 (1997).

\bibitem{PS_review}
F.~Dalfovo, S.~Giorgini, P.~Pitaevskii, Lev, and S.~Stringari,
\newblock Rev.Mod.Phys {\bf 71}, 463 (1999).

\bibitem{Pita_book}
L.~Pitaevskii and S.~Stringari,
\newblock {\em Bose-Einstein Condensation} (Oxford science publications, 2003).

\bibitem{Pethick@Smith}
C.~Pethick and H.~Smith,
\newblock {\em Bose-Einstein Condensations in Dilute Gases} (Cambridge
  University Press, 2002).

\bibitem{Bloch_colapse}
M.~Greiner, O.~Mandel, W.~H. Theodor, and I.~Bloch,
\newblock Nature {\bf 419}, 51 (2002).

\bibitem{Bloch_ex}
S.~Will {\em et~al.},
\newblock Nature {\bf 465}, 197 (2010).

\bibitem{Deepak}
D.~Iyer, R.~Mondaini, S.~Will, and M.~Rigol,
\newblock arXiv , 1408.1700v1.

\bibitem{Deepak2}
S.~Will, D.~Iyer, and M.~Rigol,
\newblock arXiv , 1406.2669v1.

\bibitem{Milburn}
G.~J. Milburn, J.~Corney, E.~M. Wright, and D.~F. Walls,
\newblock Phys. Rev. A {\bf 55}, 4318 (1997).

\bibitem{Obert_exp}
M.~Albiez {\em et~al.},
\newblock Phys. Rev. Lett. {\bf 95}, 010402 (2005).

\bibitem{jeff}
S.~Levy, E.~Lahoud, I.~Shomroni, and J.~Steinhauer,
\newblock Nature {\bf 449}, 579 (2007).

\bibitem{shin}
Y.~Shin {\em et~al.},
\newblock Phys. Rev. Lett. {\bf 92}, 050405 (2004).

\bibitem{dorond}
T.~Schumm {\em et~al.},
\newblock Nature physics {\bf 1}, 57 (2005).

\bibitem{Doron}
M.~Chuchem {\em et~al.},
\newblock Phys. Rev. A {\bf 82}, 053617 (2010).

\bibitem{Smerzi}
A.~Smerzi, S.~Fantoni, S.~Giovanazzi, and S.~R. Shenoy,
\newblock Phys.Rev.Lett {\bf 79}, 4950 (1997).

\bibitem{Spekkens&Sipe}
R.~W. Spekkens and J.~E. Sipe,
\newblock Phys. Rev. A {\bf 59}, 3868 (1999).

\bibitem{ReinhardtPRL}
D.~K. Faust and W.~P. Reinhardt,
\newblock Phys. Rev. Lett. {\bf 105}, 240404 (2010).

\bibitem{ODell}
D.~H.~J. O'Dell,
\newblock Phys. Rev. Lett. {\bf 109}, 150406 (2012).

\bibitem{ODell2}
G.~J. Krahn and D.~H.~J. O'Dell,
\newblock J. Phys. B {\bf 42}, 205501 (2009).

\bibitem{Ofir_BH}
K.~Sakmann, A.~I. Streltsov, O.~E. Alon, and L.~S. Cederbaum,
\newblock Phys. Rev. Lett. {\bf 103}, 220601 (2009).

\bibitem{Ofir_DW}
K.~Sakmann, A.~I. Streltsov, O.~E. Alon, and L.~S. Cederbaum,
\newblock Phys. Rev. A {\bf 89}, 023602 (2014).

\bibitem{OfirDW}
A.~I. Streltsov, O.~E. Alon, and L.~S. Cederbaum,
\newblock Phys. Rev. A {\bf 73}, 063626 (2006).

\bibitem{Lewenstein}
A.~Imamoglu, M.~Lewenstein, and L.~You,
\newblock Phys. Rev. Lett. {\bf 78}, 2511 (1997).

\bibitem{Wright}
E.~M. Wright, D.~F. Walls, and J.~C. Garrison,
\newblock Phys. Rev. Lett. {\bf 77}, 2158 (1996).

\bibitem{wright&walls}
E.~M. Wright, T.~Wong, M.~J. Collett, S.~M. Tan, and D.~F. Walls,
\newblock Phys. Rev. A {\bf 56}, 591 (1997).

\bibitem{castin&dalibard}
Y.~Castin and J.~Dalibard,
\newblock Phys. Rev. A {\bf 55}, 4330 (1997).

\bibitem{You}
M.~Lewenstein and L.~You,
\newblock Phys. Rev. Lett. {\bf 77}, 3489 (1996).

\bibitem{Mark}
M.~Herrera, T.~M. Antonsen, E.~Ott, and S.~Fishman,
\newblock Phys. Rev. A {\bf 86}, 023613 (2012).

\bibitem{Fisher3}
U.~R. Fischer and R.~Schutzhold,
\newblock Phys. Rev. A {\bf 78}, 061603 (2008).

\bibitem{Fisher4}
U.~R. Fischer and B.~Xiong,
\newblock Phys. Rev. A {\bf 84}, 063635 (2011).

\bibitem{of2}
D.~R. Meacher, P.~E. Meyler, I.~G. Hughes, and P.~Ewart,
\newblock J. Phys. B {\bf 24}, L63 (1991).

\bibitem{of3}
J.~A. Yeazell and C.~R. Stroud,
\newblock Phys. Rev. A. {\bf 43}, 5153 (1991).

\bibitem{Assa}
A.~Auerbach,
\newblock {\em Interacting Electrons and Quantum Magnetism} (Springer-Verlag,
  1994).

\bibitem{Kuklov4}
A.~B. Kuklov, N.~Chencinski, A.~M. Levine, W.~M. Schreiber, and J.~L. Birman,
\newblock Phys. Rev. A {\bf 55}, R3307 (1997).

\bibitem{Strunz}
L.~Simon and W.~T. Strunz,
\newblock Phys. Rev. A {\bf 86}, 053625 (2012).

\bibitem{Obert_phase}
T.~Zibold, E.~Nicklas, C.~Gross, and M.~K. Oberthaler,
\newblock Phys. Rev. Lett. {\bf 105}, 204101 (2010).

\bibitem{long}
H.~Veksler and S.~Fishman,
\newblock in preperation .

\bibitem{dorona}
E.~Boukobza, M.~Chuchem, D.~Cohen, and A.~Vardi,
\newblock Phys. Rev. Lett. {\bf 102}, 180403 (2009).

\bibitem{doronb}
E.~Boukobza, D.~Cohen, and A.~Vardi,
\newblock Phys. Rev. A {\bf 80}, 053619 (2009).

\bibitem{Tabor}
M.~Tabor,
\newblock {\em chaos and integrability in Nonlinear Dynamics} (John Wily \&
  Sons, 1989).

\bibitem{Doron_cp4}
D.~Cohen and T.~Kottos,
\newblock Phys. Rev. E {\bf 63}, 036203 (2001).

\end{thebibliography}
\end{document}